\documentclass[aps,prd,showpacs,nofootinbib]{revtex4-1}
\usepackage{latexsym}
\usepackage{amsmath,amsfonts}
\usepackage{amsbsy}
\usepackage{mathrsfs}
\usepackage{color}
\usepackage{psfrag}
\usepackage{enumerate}
\usepackage{amsmath,amssymb,calc,amsfonts}
\usepackage{latexsym}
\DeclareFontFamily{U}{rsfs}{}         
\DeclareFontShape{U}{rsfs}{m}{n}{<5> rsfs5 <6><7> rsfs7          %
  <8><9><10><10.95><12><14.4><17.28><20.74><24.88> rsfs10}{}     %
\DeclareMathAlphabet{\mathfs}{U}{rsfs}{m}{n}                     %
\definecolor{indiagreen}{rgb}{0.07, 0.53, 0.03}
\def\beq{\begin{eqnarray}}
\def\eeq{\end{eqnarray}}

\def\nn{\nonumber\\}

\def\={\stackrel{\Delta}{=}}

\def\lie{\pounds}

\def\half{{\textstyle{\frac{1}{2}}}}

\begin{document}

\title{Quasilocal rotating conformal Killing horizons}

\author{Ayan Chatterjee}\email{ayan.theory@gmail.com}
\affiliation{Department of Physics and Astronomical Science, Central University 
of Himachal Pradesh, Dharamshala -176215, India.}
\author{Avirup Ghosh}\email{avirup.ghosh@saha.ac.in}
\affiliation{Theory Division, Saha Institute of Nuclear Physics, 1/AF Bidhan 
Nagar, Kolkata 700064, INDIA.}

\begin{abstract}
The formulation of quasi-local conformal Killling horizons(CKH) is extended to include rotation. This necessitates that the
horizon be foliated by 2-spheres which may be distorted. Matter degrees of freedom which fall through the 
horizon is taken to be a real scalar field. We show that these rotating CKHs also admit a first law in differential form. 
\end{abstract}


\maketitle
\section{Introduction}

Black holes in general relativity behave like thermal objects. This analogy is based on number of
facts. It is known  that in general relativity, the surface gravity $\kappa_{H}$ of a stationary black hole must be a constant over the event horizon \cite{Bardeen:1973gs}. 
Moreover, the  first law of black hole mechanics, which  refers to 
stationary space-times admitting an event horizon and small perturbations about 
them, states that the differences in mass $M$, area $A$ and angular momentum $J$ to two nearby  stationary black hole solutions are related through $\delta M=\kappa_{H} \delta A/8\pi + \Omega_{H}\delta J.$  Additionally, according to the second law, area of black holes can never decrease in a classical process \cite{Hawking:1971vc}. Hawking's proof that due to quantum processes, black holes radiate to infinity, particles of all species at temperature $\kappa_{H}/2\pi$, implies that the laws of black hole mechanics are indeed the laws of thermodynamics \cite{Hawking:1974sw, Bekenstein:1973ur, Bekenstein:1974ax}.

These derivations, of the zeroth and the first laws of black hole mechanics, require that 
 the spacetime be stationary (The derivation of the second law however does not require that the spacetime is stationary
 but uses the teleological notion of an event horizon). In that case, the future event horizon of a stationary black hole is 
 a Killing horizon. However,  not all Killing horizons require that the entire spacetime be stationary. Indeed, one may have 
 Killing horizons which has a 
timelike Killing vector field in the neighbourhood of the horizon only. Since Killing horizons give 
a local description of black hole horizons, one may enquire if the laws of black hole mechanics hold 
good for Killing horizons too. Remarkably, the laws of black hole mechanics hold good for bifurcate Killing 
horizons. The framework of Killing Horizon is also useful to  study and unravel the origin of entropy 
and black hole thermodynamics \cite{Wald:1995yp, Wald:1993nt, Iyer:1994ys, Jacobson:1993vj, Youm:1997hw, 
Carlip:1999cy, Dreyer:2013noa}. Killing horizons are not the only local description of black hole
boundaries, one can construct more. The notion of trapping horizons is one such description \cite{Hayward:1993wb, Hayward:1994yy}. The 
formalism of isolated horizons and dynamical horizons, which are related to trapping horizons,
\cite{Ashtekar:1998sp, Ashtekar:2000sz, Ashtekar:2000hw, Ashtekar:2001is, 
Ashtekar:2001jb, Chatterjee:2006vy, Chatterjee:2008if, Ashtekar:2002ag, Ashtekar:2003hk, Ashtekar:2004cn, 
Booth:2003ji, Booth:2006bn, Booth:2007wu} 
have been used to address some questions regarding classical and quantum mechanics of black holes. Within
this framework of isolated and dynamical horizons, one can establish the laws of black hole mechanics and
show the entropy of black holes can be determined by counting the black hole microstates residing on the
horizon only \cite{Smolin:1995vq, Krasnov:1996tb, 
Rovelli:1996dv, Ashtekar:1997yu, Ashtekar:1999wa, Ghosh:2006ph, Ghosh:2008jc, 
Ghosh:2011fc, 
Ghosh:2013iwa}.

Another class of horizons that are of interest are conformal Killing 
horizons (CKH) \cite{DyerHonig,Suldyer, Sultana:2005tp, 
Jacobson:1993pf, Nielsen:2012xu}. These horizons capture the essence of dynamical situations. CKHs are 
null hypersurfaces whose null geodesics are orbits of a conformal Killing field. More precisely, if $\xi^{a}$ is 
a vector field satisfying $\lie_{\xi}g_{ab}=2fg_{ab}$,
and is null, it generates a CKH for the metric $g_{ab}$. Since $\xi^{a}$ generates a 
null surface, and generates geodesics, one can define an acceleration given by the relation
$\xi^{b}\nabla_{b}\xi^{a}=\kappa_{\xi} \xi^{a}$. Then, it arises that the quantity 
$(\kappa_{\xi}-2f)$ which is a combination of the acceleration 
of the conformal Killing vector and the conformal factor, is Lie dragged along 
the horizon. Moreover,  if the stress-energy tensor satisfies the strong energy condition, then this 
quantity is a constant on the horizon. It can therefore be interpreted as a temperature. Thus, a form of zeroth 
law holds for these horizons.The
existence of such a law is not very surprising given the fact that these horizons are generated by conformal
Killing vectors. Conformal Killing vector fields, just like that of a Killing vector, provide a sense 
\emph{time}. Consider for example a horizon
generated by Killing vector. In this example, the zeroth law holds and the surface gravity is the acceleration generated by 
the Killing vector. The situation is similar for CKH only that the quantity which remains constant on the surface of the 
horizon cross- section includes the conformal factor. One may enquire if a quasi-local formulation of conformal Killing horizons may be
developed. This extension would be similar to the generalisation of Killing horizons to isolated horizons. A 
conformal Killing horizon is defined retroactively and one needs to know the full space-time history and a globally 
defined conformal Killing 
vector. In contrast, a quasi-local conformal Killing horizon only requires the existence of a null hypersurface 
generating vector field.\footnote{One can construct solutions of Einstein's equations for gravity and matter which admit 
a conformal Killing horizon \cite{Sultana:2005tp}. It may also be possible to construct solutions 
admitting a quasi-local conformal horizon.} Indeed, it has been shown that it is possible to broaden 
the boundary conditions to construct a quasi-local conformal Killing horizon and that these horizons have a 
zeroth law and a first law \cite{Chatterjee:2014jda}. The basic idea goes as follows: Consider a
spacetime $\mathcal{M}$ having a null boundary $\Delta$ with 
non- zero expansion ($\theta =-2\rho\neq 0$) but the null generators of $\Delta$ 
are assumed to be shear-free. These 
conditions guarantee that the null generators $l^{a}$ are conformal 
Killing vectors on $\Delta$. Clearly,  these null surfaces are not expansion 
free, they may be growing. Indeed, $\lie_l ~{}^{2}\epsilon=\theta~{}^{2}\epsilon$ and hence,
they are good candidates for growing horizons. It was further assumed in \cite{Chatterjee:2014jda} that 
the horizons expand due to the reason that matter fields fall through these horizons. For definiteness, this
matter field was taken to be a massless scalar field satisfying the condition
 $\lie_l\varphi\stackrel{\Delta}{=}-2\rho\,\varphi$. This 
assumption is motivated by the fact that $l^a$ is a conformal Killing vector on $\Delta$. The first law for quasi- local CKHs was shown to get the form $dU=TdS$ along with a flux term arising due to matter fields falling through the horizon. 

Even if the problem we are dealing with is concerned with dynamical evolution of black hole horizons, we argue that this result, that a differential first law exists, is not surprising. In a fully dynamical situation, a black hole is far from equilibrium and thermodynamic quantities like temperature cannot be defined. This gets reflected in the first law for dynamical black holes. For example, in the case of dynamical horizons \cite{Ashtekar:2003hk,Ashtekar:2004cn}, the surface gravity cannot be pulled out of the integrals on the horizon cross sections and hence the first law can not be written in a differential form. However, we do not consider such an extreme case but concentrate on a simpler situation where the the horizon is generated by a conformal Killing vector field. This helps us to address dynamical black holes but in a simpler setting where the horizon has a preferred sense of “time”. The quantity quantity $\kappa+2\rho$ can be shown to be constant on $\Delta$ and can be interpreted as a temperature (note that if a system is in thermodynamic equillibrium then one can consistently define a notion of temperature. However the converse in not true i.e having a notion of a temperature does not imply that the system is in equillibrium). It then turns out that the first law may be written in a differential form for the CKH.

However, since the most useful application of these geometrical structures are in the 
dynamical evolution of black holes, one must address some further issues left out in \cite{Chatterjee:2014jda}. Suppose that the horizon $\Delta$ is rotating with some angular momentum $J$. What are the boundary conditions which will ensure a zeroth law? Are these boundary conditions enough to construct the space of solutions of general relativity? How would one define an angular momentum? By how much do the first law change in the presence of rotation ? Can the first law be written in a differential form ? In this paper, we answer these questions.

The plan of the paper is as follows. We start by developing the geometry of a rotating
quasi-local conformal Killing horizon. To account for rotation, we assume that there is a spacelike conformal Killing vector $\phi^a$ on $\Delta$ such that it commutes with $l^a$. We show that the zeroth law is valid. Using the first order action,
we construct the space of solutions of Einstein's theory and show that a well- defined hypersurface independent symplectic structure exits. In the next section, we construct angular momentum as a Hamiltonian corresponding to the axial conformal Killing vector field. We derive the first law for rotating horizons and show that it may be written in a differential form.
We will closely follow the formalism already laid down in \cite{Chatterjee:2014jda}. However, there are some crucial changes since we allow for rotating cross-sections.

\section{Boundary conditions for rotating quasilocal CKH}

Let $\mathcal{M}$ be a $4$- manifold equipped
with a metric $g_{ab}$ of signature $(-,+,+,+)$. We assume that all fields on $\mathcal{M}$ are smooth. Let 
$\Delta$ be a null hypersurface of $\mathcal M$. On this hypersurface, we construct the Newman- Penrose basis $(l,\, n\, m\, \bar{m})$,
where $l^a$ is the future directed null normal and $n^{a}$ the transverse and future directed null vector field to $\Delta$.
The set of complex null vector field $(m,\,\bar{m})$ are taken to be tangential to $\Delta$.  
This null tetrad $(l,\, n\, m\, \bar{m})$ satisfy the condition $l.n=-1=-m.\bar{m}$, while 
all other scalar products 
vanish. The degenerate metric on this hypersurface $\Delta$ is denoted by $q_{ab}$.
The expansion $\theta_{l}$ of the null normal is defined by
$q^{ab}\nabla_{a} l_b$. In terms of the Newman- Penrose formalism,
$\theta_l=-2\rho$ (see appendix A of \cite{Chatterjee:2014jda} or \cite{Chandrasekhar:1985kt} for details). The accelaration of $l^a$ can be obtained from the 
expression $l^a\nabla_a~l_b=(\epsilon+\bar\epsilon)l_b$ and is given by 
$\kappa_l :=(\epsilon+\bar\epsilon)$.\footnote{To avoid cumbersome notation, we will do 
away with the subscripts $(l)$ from now on if no confusion arises.} One may further define an equivalance class of 
null normals $[l^{a}]$ such that $l$ and $l'$ belong to the same equivalance 
class if $l'=cl$ where $c$ is a constant on $\Delta$.\\


{\it{Definition}}: A null hypersurface $\Delta$ of $\mathcal{M}$ will be called 
quasi-local conformal horizon if the following conditions hold.:
\begin{enumerate}
\item $\Delta$ is topologically  $S^2\times R$ and null.
\item The shear $\sigma$ of $l^{a}$ vanishes on $\Delta$ for any null normal $l^{a}$.
\item All equations of motion hold at $\Delta$ and the stress-
energy tensor $T_{ab}$ satisfies the null energy condition on $\Delta$.
\item If $\varphi$ is a matter field then it must satisfy 
$\lie_l\varphi=-2\rho\,\varphi$ on $\Delta$ for all null normals $l^{a}$.
\item  The quantity $\left[2\rho+\epsilon + \bar{\epsilon}\right]$ is lie 
dragged for any null-normal $l^{a}$.

\item There is a spacelike axial conformal Killing vector $\phi^a$ on $S_\Delta$ such that $\lie_\phi q_{ab}=-2g~q_{ab}$ and it commutes with the $l^a$ viz. $[\phi,l]^a=0$
\item $\phi^a$ has closed circular orbits of length $2\pi$ and vanishes on exactly two generators of $\Delta$.
\end{enumerate}


The conditions $1- 5$ are already described in \cite{Chatterjee:2014jda} where it has been established that they describe a quasi- local conformal horizon. To include rotations, we must take into account a description of the vector fields that generate the 2- spheres of the horizons and incorporate the boundary conditions on these fields. Thus, for a rotating CKH, apart from the above conditions 
$(1- 5)$, the conditions $(6 -8)$ are also assumed to be true.
The first condition imposes restrictions on the topology of the hypersurface. The cross- sections of such quasi-local horizons may admit other topologies but we do not include such generalities here.

The second boundary condition concerns shear which measures 
the amount of gravitational flux 
flowing across the surface.  We assume that the gravity flux be vanishing. This 
boundary condition on 
the shear $\sigma$ of null normal $l^{a}$ has several consequences. First, 
since $l_a$ is hypersurface orthogonal, the Frobenius theorem implies that 
$\rho$ is real and $\kappa=0$. Secondly, $l^{a}$ is twist- free and a geodetic vector field. 
The acceleration of $l^a$ is given through expression $l^a\nabla_a~l^b=(\epsilon+\bar\epsilon)l^b$ where,  $\kappa_l :=(\epsilon+\bar\epsilon)$. 
The acceleration varies in the equivalence class $[cl^{a}]$ since in the absence of the knowledge of  asymptotics, the acceleration cannot be fixed. Thirdly, a Ricci identity is given by 
\beq
D\sigma-\delta\kappa=\sigma(\rho+\bar{\rho}+3\epsilon-\bar{\epsilon}
)-\kappa(\tau-\bar{\pi}+\bar{\alpha}+3\beta)+\Psi_0,
\eeq
where $D=l^{a}\nabla_{a}$, $\delta=m^{a}\nabla_{a}$, $\Psi_{0}$
is one of the Weyl scalars and the other quantities are the Newman- Penrose 
scalars (see \cite{Chandrasekhar:1985kt}
for details). For $\sigma\stackrel{\Delta}{=}0$, it implies $\Psi_0\stackrel{\Delta}{=}0$.
Further, it can be seen that the null normal $l^{a}$ is such that
\beq
\underleftarrow{\nabla_{(a}l_{b)}}\stackrel{\Delta}{=}-2\rho\, 
m_{(a}\bar{m}_{b)}
\eeq
which implies that $l^a$ is a conformal Killing vector on $\Delta$. Moreover, 
the Raychaudhuri equation implies that
$R_{ab}l^{a}l^{b}\neq 0$ and hence $-R^{a}{}_{b}l^{b}$ can have components 
which are tangential as well as transverse to $\Delta$.

The third boundary condition only implies that the field equations of gravity 
be satisfied and the matter fields be such
that the stress tensor satisfies the null energy condition. In the fourth
condition, we 
have kept open the possibility that matter fields may
cross the horizon and the horizon may grow. The matter field is taken to be a 
massless scalar field $\varphi$ which behaves in a certain way
which mimics it's conformal nature. The fifth condition can also be motivated if the fact that
$\left(2\rho+\epsilon+\bar{\epsilon}\right)$ remains invariant
under conformal transformations \cite{Jacobson:1993pf, Sultana:2005tp}. This can
also be shown as follows. A conformal 
transformation of the metric amounts to a conformal transformation of the 
two-metric on $\Delta$.
Under a conformal transformation $g_{ab}\rightarrow \Omega^2\,g_{ab}$ one needs a new covariant derivative operator
which annihilates the conformally transformed metric. Under such a conformal transformation $l^a\rightarrow l^a,l_a\rightarrow \Omega^2 l_a,n^a\rightarrow \Omega^{-2} n^a,n_a\rightarrow n_a,m^a\rightarrow \Omega^{-1}m^a,m_a\rightarrow \Omega m_a$. The new derivative 
operator is such that it transforms as
\begin{equation}
\nabla_al_b\rightarrow\Omega^2\nabla_al_b+2\Omega\partial_a\Omega~l_b-\Omega^2\left[l_c\delta^c_a\partial_b\log{\Omega}+l_c
\delta^c_b\partial_a\log{\Omega}-g_{ab}g^{cd}l_c\partial_d\log{\Omega}\right]
\end{equation}
If one defines a one- form as $\omega_a\=-n^b\underleftarrow{\nabla_a}l_b$, it 
transforms under the conformal transformation
as
\begin{equation}\label{conftr_omega}
\tilde{\omega}_a\=\omega_a+2\partial_a\log{\Omega}-\partial_a\log{\Omega}-n_al^c\partial_c\log{\Omega}
\end{equation}
It follows that the Newman- Penrose scalars $\rho=-m^a\bar{m}^b\nabla_al_b$ and 
$\sigma=-\bar{m}^a\bar{m}^b\nabla_al_b$ transform in such a way that 
$(2\rho + \epsilon+\bar\epsilon )$ remains invariant under a conformal 
transformation.  Further, since the Weyl tensor is invariant under a conformal rescaling, it follows that $\Psi_1\stackrel{\Delta}{=}0$ in this case.

The sixth and the eighth conditions on the vector field $\phi^{a}$ are motivated from the similar conditions on $l^{a}$ and that the vector field must preserve the geometric structures on 
$\Delta$. The seventh condition is just the statement that the integral curves of $\phi^{a}$ cover the sphere.

\subsection*{Gauge choices}
Since the null tetrad is typically not a coordinate basis, it leads to non-trivial commutation relations\cite{badri}. The following two commutation relations are useful to choose a gauge so as to make the commputations simpler.
\beq
  (\delta D-D\delta)f &=& (\bar{\alpha}+\beta-\bar\pi)Df + \kappa\Delta
  f - (\bar{\rho}+\epsilon-\bar{\epsilon})\delta f
  -\sigma\bar{\delta}f\,,\\
  (\bar{\delta}\delta-\delta\bar{\delta})f &=& (\bar{\mu}-\mu)Df +
  (\bar{\rho}-\rho)\Delta f + (\alpha-\bar{\beta})\delta f -
  (\bar{\alpha}-\beta)\bar{\delta}f\,,
\eeq
where $D=l^a\nabla_a, \Delta=n^a\nabla_a, \delta=m^a\nabla_a$. Since $l^a$ is geodetic one can choose a coordinate $v$ on $\Delta$ such that $\lie_lv=1$. If we choose the function $f=v$,then from the above commutation relations it follows that $\mu=\bar\mu$ and $\pi=\alpha+\bar\beta$.

\subsection*{Calculation of $d\omega$}
Since we deal with rotating horizons it is useful to obtain an expression for $d\omega$. This is because $d\omega$ contains the imaginary part of the Weyl scalar $\Psi_2$. The information of angular momentum is contained in this scalar. In the case of an isolated horizon $d\omega_{IH}=2(Im\Psi_2)~^2\epsilon$. However unlike an isolated horizon the induced connection $(\omega)$ on a CKH is not Lie-dragged along the horizon but is however related to $\omega_{IH}$ through a conformal transformation as demonstrated in eq \eqref{conftr_omega}. So one needs to check if the induced connection on a CKH also contains the information of angular momentum. Further it would also be a check if the results of isolated horizon can be recovered if one goes from $\omega$ to $\omega_{IH}$ via a conformal transformation \eqref{conftr_omega}. We start with the definition of Riemann tensor given as $\left[\nabla_a\nabla_b-\nabla_b\nabla_a\right]X^c=-R_{abd}~{^c}X^d$, for a vector field $X^a$. Putting $X^a=l^a$, we get,
\beq
\left[\nabla_a\nabla_b-\nabla_b\nabla_a\right]l^c=-R_{abd}~{^c}l^d
\eeq

Consider the left hand side of the above equation. Using the expression for $\nabla_al_b$ given in \cite{Chatterjee:2014jda} we get,
\beq
\nabla_a(\nabla_bl^c)-\nabla_b(\nabla_al^c)&=&\nabla_a\left(\omega_b l^c-\bar\rho \bar m_bm^c-\rho m_b\bar m^c\right)-(a\leftrightarrow b)\nn
&=&\nabla_a\left(\omega_b\right) l^c+\omega_b\left(\omega_a l^c-\bar\rho \bar m_bm^c-\rho m_b\bar m^c\right)-\nabla_a\left(\bar\rho \bar m_bm^c-\rho m_b\bar m^c\right)-(a\leftrightarrow b)\nn
\eeq

Contracting the above by $n_c$, pulling back the expression on to $\Delta$ gives and using the gauge choices, gives,
\beq
\left(\underleftarrow{\left[\nabla_a\nabla_b-\nabla_b\nabla_a\right]}l^c\right)n_c&\stackrel{\Delta}{=}&-\underleftarrow{\nabla_a\omega_b}-\left(\bar\rho \bar m_b\left(\bar{\pi}n_a-\bar{\lambda}\bar{m}_a-\bar{\mu} m_a\right)+\rho m_b\left({\pi}n_a-{\lambda}{m}_a-{\mu} \bar m_a\right)\right)-(a\leftrightarrow b)\nn
&=&-2\underleftarrow{\nabla_{[a}\omega_{b]}}-2\rho\left(\bar{\pi}\bar m_{[b}n_{a]}+{\pi}m_{[b}n_{a]}\right)
\eeq
Now the Weyl tensor can be written in terms of the curvatures as follows:
\beq
C_{abcd}&=&R_{abcd}-\left(g_{a[c}R_{d]b}-g_{b[c}R_{d]a}\right)+\frac{1}{3}Rg_{a[c}g_{d]b}
\eeq
We also expand the Ricci tensor and the Weyl tesor in a Newman-Penrose basis and obtain the following results:
\beq
R_{ab}&=&2\Phi_{00}n_an_b+2\Phi_{22}l_al_b+2\Phi_{02}\bar m_a\bar m_b+2\Phi_{20}m_am_b\nn
&&~~+\frac{1}{2}(4\Phi_{11}-12\Lambda)(n_al_b+l_an_b)+\frac{1}{2}(4\Phi_{11}+12\Lambda)(m_a\bar m_b+m_b\bar m_a)\nn
&&~~~~~~~~-2\Phi_{01}(n_a\bar m_b+n_b\bar m_a)-2\Phi_{10}(n_am_b+n_bm_a)-2\Phi_{21}(l_a\bar m_b+l_b\bar m_a)-2\Phi_{12}(l_am_b+l_bm_a)\\
\nn
C_{abcd}l^cn^d&=&4(Re[{\Psi}_2])l_{[a}n_{b]}+2{\Psi}_3l_{[a}m_{b]}+2\bar{\Psi}_3l_{[a}\bar m_{b]}-2\bar{\Psi}_1n_{[a}\bar m_{b]}-2{\Psi}_1n_{[a}m_{b]}+4i(Im[{\Psi}_2])m_{[a}\bar m_{b]}
\eeq

Combining the above expressions one gets,

\beq
\left[\underleftarrow{\left(\nabla_a\nabla_b-\nabla_b\nabla_a\right)}l^c\right]n_c&=&-R_{\underleftarrow{ab}d}~{^c}l^dn_c=-R_{\underleftarrow{ab}cd}l^cn^d=-C_{\underleftarrow{ab}cd}l^cn^d+2\Phi_{01}n_{[a}\bar m_{b]}+2\Phi_{10}n_{[a}m_{b]}\\
-2\underleftarrow{\nabla_{[a}\omega_{b]}}-2\rho\left(\bar{\pi}n_{[a}\bar m_{b]}+{\pi}n_{[a}m_{b]}\right)&=&-4i(Im \Psi_2)m_{[a}\bar m_{b]}+2\Phi_{01}n_{[a}\bar m_{b]}+2\Phi_{10}n_{[a}m_{b]}
\eeq
Using the gauge choices and the Ricci identity $m^a\nabla_a\rho=\rho(\bar\alpha+\beta)+\Phi_{01}$, 
\beq
-2\underleftarrow{\nabla_{[a}\omega_{b]}}&=&-4i(Im \Psi_2)m_{[a}\bar m_{b]}+(2\rho(\bar\alpha+\beta)2\Phi_{01})n_{[a}\bar m_{b]}+(2\rho(\alpha+\bar\beta)2\Phi_{10})n_{[a}m_{b]}\nn
&=&-4i(Im \Psi_2)m_{[a}\bar m_{b]}+(2m^a\nabla_a\rho)n_{[a}\bar m_{b]}+(2\bar m^a\nabla_a\rho)n_{[a}m_{b]}
\eeq
Which immediately implies that,
\beq
{d\omega}&=&2i(Im \Psi_2)m\wedge\bar m-n\wedge d\rho
\eeq
Now we can again rewrite this expression in terms of the conformally transformed connection $\tilde\omega$ with the choice $\lie_l\log\Omega=\rho$ to get,
\beq
{d\tilde\omega}&=&2(Im \Psi_2)~^2\epsilon=2(Im \tilde\Psi_2)~^2\tilde\epsilon.
\eeq
which is exactly the espression obtained for an isolated horizon.

\subsection*{Symmetries of Quasi-local conformal horizons}
In this section we discuss the infinitesimal symmetries of a quasi-local conformal horizon. In the absence of asymptotic infinity, symmetry in our case would mean preserving the relevant geometric structures of the horizon. It is at once clear that only vector fields tangent to $\Delta$ preserve the boundary conditions. Let us consider a general case first. Since $\Delta$ is a hypersurface it follows that the vector fields tangent to $\Delta$ form a closed Lie algebra. Let $\xi^{a}$ be a vector field tangent to the horizon. Then $\xi^{a}$ will be said to be a symmetry generating vector field if the following conditions hold
\begin{enumerate}
\item It preserves the equivalence class of null normals i.e 
\beq
[\xi,l]^a\stackrel{\Delta}{=}cl^a \text{~~~~~where $c$ is a constant on $\Delta$}
\eeq
\item $\xi^{a}$ is a conformal Killing vector on $\Delta$. If $q_{ab}$ is the degenerate metric on and $h$ is smooth function on $\Delta$ then, 
\beq
\lie_\xi q_{ab}\stackrel{\Delta}{=}h~q_{ab}
\eeq
\item It Lie drags the conformally transformed connection $\tilde\omega$ on $\Delta$ with the conformal factor satisfying $\lie_l\log\Omega\stackrel{\Delta}{=}\rho$.
\beq
\lie_\xi\tilde\omega\stackrel{\Delta}{=}0
\eeq
\end{enumerate}
Note that the third condition $\lie_\phi\tilde\omega$ is analogous to the one for a weakly isolated horizon \cite{Ashtekar:2001is}. It is immediately clear that $\phi^a$ satisfies the first two conditions. However, to qualify as a symmetry vector field, the third condition on $\phi^{a}$ must also be met. It implies certain conditions on fields. First, we note the following:
\beq
\lie_l~^2\epsilon=-2\rho~^2\epsilon~~~~~\lie_\phi~^2\epsilon=-2g~^2\epsilon
\eeq
Together the above two equations can be rewritten together as,
\beq
[\lie_l,\lie_\phi]~^2\epsilon=-2(\lie_lg-\lie_\phi\rho)~^2\epsilon
\eeq
Since $l^a$ commutes with $\phi$ it follows that $\lie_lg=\lie_\phi\rho$. The symmetry vector field puts the following restriction on the conformal factor $\Omega$,  given by $\lie_l\log\Omega=\rho $.
It immediately follows that,
\beq\label{fixomega}
\lie_l(\lie_\phi\log\Omega-g)&=&0
\eeq
Since $\log\Omega$ is not completely determined by the condition $\lie_l\log\Omega=\rho$ one still has the liberty of choosing the spatial dependendence of $\log\Omega$ such that $\lie_\phi\log\Omega=g$. Then condition $\eqref{fixomega}$ ensures that it holds everywhere on $\Delta$. The above calculation shows that one can choose $\log\Omega$ such that both $l^a$ and $\phi^a$ are Killing vectors of the conformally transformed metric.

\section{Action principle, phase space and the first law}
We are interested in constructing the space of solutions of general relativity, 
and we
use the first order formalism in terms of tetrads and connections to construct the covariant phase- space. 
For the first order theory, we take the fields on the manifold to be
($e_{a}{}^{I},\, A_{aI}{}^{J},\, \varphi$), where $e_{a}{}^{I}$ is the co- 
tetrad, $A_{aI}{}^{J}$
is the gravitational connection and $\varphi$ is the scalar field.
The Palatini action in first order gravity with a scalar field is given by:
\begin{equation}\label{lagrangian1}
S_{G+M}=-\frac{1}{16\pi G}\int_{\mathcal{M}}\left(\Sigma^{I\!J}\wedge 
F_{I\!J}\right)-\frac{1}{2}\int_{\mathcal{M}}d\varphi\wedge {}{\star} 
d\varphi\;
\end{equation}
where $\Sigma^{IJ}=\half\,\epsilon^{IJ}{}_{KL}e^K\wedge e^L$, $A_{IJ}$ is a 
Lorentz $SO(3,1)$ connection 
and $F_{IJ}$ is a curvature two-form corresponding to the connection given by
$F_{IJ}=dA_{IJ}+A_{IK}\wedge A^{K}~_{J}$. The action might have to be 
supplemented with boundary terms to 
make the variation well defined.
We now check that the variational principle is well- defined if the boundary 
conditions on the fields, as given in the previous section, hold.

The Lagrangian $4$- form for the fields ($e_{a}{}^{I},\, A_{aI}{}^{J},\, 
\varphi$) is given in the following way.
\begin{equation}
L_{G+M}=-\frac{1}{16\pi G}\left(\Sigma^{I\!J}\wedge 
F_{I\!J}\right)-\frac{1}{2}d\varphi\wedge \star d\varphi .
\end{equation}
The first variation of the action leads to equations of motion and boundary 
terms. The equations of motion consist of the following equations. 
First, variation of the action with respect to the connection implies that the 
curvature $F^{IJ}$ is related to the 
Riemann tensor $R^{cd}$,  through the relation 
$F_{ab}{}^{IJ}=R_{ab}{}^{cd}\,e^{I}_{c}e^{J}_{d}$. Second, variation with 
respect 
to the tetrads lead to the Einstein equations and third, the first variation of 
the matter 
field gives the equation of motion of the matter field (details can be found in 
the appendix of \cite{Chatterjee:2014jda}).  
On- shell, the first variation is given by the following boundary terms
\begin{equation}\label{var_Lagrangian}
\delta L_{G+M} := d\Theta(\delta)=-\frac{1}{16\pi 
G}d\left(\Sigma^{IJ}\wedge\delta A_{IJ}\right)-d(\delta\varphi\star d\varphi).
\end{equation}
The quantity $\Theta(\delta)$ is called the symplectic potential and the boundary terms are to be evaluated on the initial and final spacelike boundaries $M_{-}$, $M_{+}$, asymptotic infinity and the internal boundary $\Delta$. However, since fields are 
set fixed on the initial and the final hypersurfaces, they vanish. The boundary 
conditions at infinity are
assumed to be appropriately chosen and they can be suitably taken care of. The 
only terms which are of
relevance for this case are the terms on the internal boundary. 
On the conformal horizon $\Delta$, the $\Sigma^{IJ}$ is given by \cite{Chatterjee:2014jda}
\begin{equation}\label{sigma_exp} 
\underleftarrow{\Sigma}^{IJ}\=2l^{[I}n^{J]}~^{2}\epsilon+2n\wedge(im~l^{[I}\bar{
m}^{J]}-i\bar{m}~l^{[I}m^{J]}),
\end{equation}
and the connection is given by \cite{Chatterjee:2014jda} 
\begin{eqnarray}\label{connection_delta}
\underleftarrow{A_{a}{}_{IJ}}&\stackrel{\Delta}{=}&
2\left[(\epsilon+\bar{\epsilon})n_a 
-(\bar{\alpha}+\beta)\bar{m}_a-(\alpha+\bar{\beta})m_a\right]\, 
l_{[I}n_{J]}+2(-\bar{\kappa}n_a 
+\bar{\rho}\bar{m}_a)\, m_{[I}n_{J]}+2(-{\kappa}n_a 
+{\rho}{m}_a)\,\bar{m}_{[I}n_{J]}\nn
&+& 2(\pi n_a+-\mu\bar{m}_a-\lambda m_a)\, m_{[I}l_{J]}+2(\bar{\pi} n_a 
-\bar{\mu}{m}_a-\bar{\lambda}\bar{m}_a)\,\bar{m}_{[I}l_{J]}\nn
&+& 2\left[-(\epsilon-\bar{\epsilon})n_a + 
(\alpha-\bar{\beta}) m_a+(\beta-\bar{\alpha})\bar{m}_a \right]\, m_{[I}\bar{m}_{J]}.
\end{eqnarray}
%

%
Consider the gravity terms first\footnote{In our case it might not be possible to define a unique covariant derivative on $\Delta$. However, since in the the calculations $l^a\nabla_a$ acts only on functions, the amibiguity do not play a role.} By using the Ricci identities
in terms of Newman-Penrose coeffecients, 
\beq\label{NP_eqna}
D\rho&=&\rho^2+\rho(\epsilon+\bar{\epsilon})+\Phi_{00}\nonumber\\
m^a\nabla_a\rho&=&\rho(\bar\alpha+\beta)+\Phi_{01},
\eeq
we find from equations\eqref{var_Lagrangian}, \eqref{sigma_exp} and \eqref{connection_delta} 
that
\beq
\Sigma^{IJ}\wedge\delta A_{IJ}&=&-2~^2\epsilon\wedge\delta[(\epsilon+\bar{\epsilon})n-(\alpha+\bar\beta)m-(\bar\alpha+\beta)\bar m]+2(n\wedge im)\wedge\delta(\rho\bar{m})-2(n\wedge 
i\bar{m})\wedge\delta(\rho{m})\nn
&=&-2~^2\epsilon\wedge\delta\left[\left(\frac{D\rho}{\rho}-\rho-\frac{\Phi_{00}}
{\rho}\right)n-\left(\frac{\bar m^a\nabla_a\rho}{\rho}-\frac{\Phi_{01}}{\rho}\right)m-\left(\frac{m^a\nabla_a\rho}{\rho}-\frac{\Phi_{10}}{\rho}\right)\bar m\right]\nn
\nn
&&~~~~~~~~~~~~~~~~~~~~~+2(n\wedge im)\wedge\delta(\rho\bar{m})-2(n\wedge 
i\bar{m})\wedge\delta(\rho{m})\nn
&=&d\left[2~^2\epsilon~\delta(\log{\rho})\right]-4n\wedge 
{}^2\epsilon\,\delta\rho+2~^2\epsilon\wedge\delta\left[\left(\rho+\frac{\Phi_{00
}}{\rho}\right)n-\left(\frac{\Phi_{10}}{\rho}\right)m-\left(\frac{\Phi_{01}}{\rho}\right)\bar m\right]\nn
&&~~~~~~~~~~~~~~~~~~~~~~+4n\wedge^2\epsilon~\delta\rho+2\rho 
n\wedge\delta~^2\epsilon\nn
\nn
&=&d\left[2~^2\epsilon~\delta(\log{\rho})\right]+2~^2\epsilon\wedge\delta\left[
\left(\frac{{R}_{11}}{2\rho}\right)n-\left(\frac{{ R}_{14}}{2\rho}\right)m-\left(\frac{{ R}_{13}}{2\rho}\right)\bar m\right]+\delta\left(2\rho 
n\wedge~^2\epsilon\right)
\eeq

Note that this expression is analogous to the one in \cite{Chatterjee:2014jda} but have few differences. Here, we have terms like $~^2\epsilon\wedge\delta[(\alpha+\bar\beta)m+(\bar\alpha+\beta)\bar m]$ were ignored in \cite{Chatterjee:2014jda} owing to the fact that the horizon 2-surfaces were assumed to be round 2- spheres. Here, we have to take them into account and hence these terms have been retained throughout. The matter Lagrangian leads to the following variation:
\beq
(\delta\varphi\star d\varphi)
&=&-d\left(\frac{1}{2}\delta\varphi^2~^2\epsilon\right)+\delta\left(\varphi~d\varphi\right)\wedge~^2\epsilon\nn
&=&-d\left(\frac{1}{2}\delta\varphi^2~^2\epsilon\right)-\frac{1}{2}\delta\left(\frac{{ T}_{11}}{\rho}n-\frac{{ T}_{14}}{\rho}m-\frac{{ T}_{13}}{\rho}\bar m\right)\wedge~^2\epsilon
\eeq
Adding everything up, one finds that a total $\delta$ term survives and can be written as:
\beq
\int_\Delta\Theta(\delta)=-\frac{1}{16\pi G}\int_\Delta\delta\left(2\rho 
n\wedge~^2\epsilon\right)
\eeq
Since Einstein's equations give $R_{11}=8\pi G \, T_{11}, R_{13}=8\pi G \, T_{13}$ and $R_{14}=8\pi G \, T_{14}$ only $-\left(2\rho\, n\wedge\,^2\epsilon\right)$ survives. Note the the total derivative terms contribute only at the two-surfaces at the intersection of $\Delta$ with  $M_{-}$, $M_{+}$. However since the fields are fixed at $M_{-}$, $M_{+}$ their contribution is zero. Thus, if one adds the term $16\pi G\,S^{'}= \int_{\Delta}\left(2\rho\, n\wedge\,^2\epsilon\right)$ to the action, 
it is well defined for the set of boundary conditions on $\Delta$. As we shall see below, since this is a boundary term, it does not contribute to the symplectic structure.

\subsection*{Covariant phase- space and the symplectic structure}
 The symplectic potential $\Theta (\delta)$ in eqn. \eqref{var_Lagrangian} is a $3$-form in space-time and a $0$-form in phase space.
Given the symplectic potential, one can construct the symplectic current 
$J(\delta_1,\delta_2)= \delta_1\Theta(\delta_2)-\delta_2\Theta(\delta_1)$, 
%
\begin{equation}
\Omega(\delta_{1}, \,\delta_{2})= \int_MJ-\int_{S_B}j
\end{equation}
where $S_B$ is the 2-surface at the intersection of any hypersurface  $M$ with 
the boundary B. The quantity $j(\delta_1,\delta_2)$ is called the boundary symplectic current. 
In this case, the boundaries are the null boundary $\Delta$ and a boundary at infinity. We shall assume that the fall- off on the fields are such that the integrals over the surface at infinity vanishe. Indeed, if the asymptotic is flat the fall- off on the fields can be set such that the contribution from the cylinder at infinity is zero. Thus, the contributions to the symplectic structure would come from the spheres on the inner boundary $\Delta$.

Our strategy shall be to construct the symplectic structure for the action 
given in eqn. \eqref{lagrangian1}. Let us first look 
at the Lagrangian for gravity. The symplectic potential in this case is 
given by, $16\pi G\Theta(\delta)=-\Sigma^{I\!J}\wedge \delta A_{I\!J}$. The 
symplectic current 
is therefore given by,
\begin{equation}\label{symplectic_current1}
J_G(\delta_1,\delta_2)=-\frac{1}{8\pi 
G}\,\delta_{[1}\Sigma^{IJ}\wedge~\delta_{2]}A_{IJ}
\end{equation}
The above expression eqn. \eqref{symplectic_current1}, when pulled back and 
rescticted to the surface
$\Delta$ gives
\beq\label{symplec_pulled_back}
\underleftarrow{J_G}(\delta_1,\delta_2) 
&\stackrel{\Delta}{=}&-2\,\delta_{[1}~^2{\epsilon}
\wedge\delta_{2]}\left\{(\epsilon+\bar{\epsilon})n-(\alpha+\bar{\beta}
)m-(\bar{\alpha}+\beta)\bar{m}\right\}+2\,\delta_{[1}(n\wedge im)\wedge\delta_{2]}(\bar{\rho}\bar{m})
-2\,\delta_{[1}(n\wedge i\bar{m})\wedge\delta_{2]}(\rho m) \nonumber\\
&\stackrel{\Delta}{=}&-\frac{1}{4\pi G}
\left[d\left(\delta_{[1}~^2{\epsilon}~\delta_{2]}\log{\rho}\right)+\delta_{[1
}~^2{\epsilon}\wedge\delta_{2]}
\left\{\left(\frac{\Phi_{00}}{\rho}
\right)n-\left(\frac{\Phi_{10}}{\rho}
\right)m-\left(\frac{\Phi_{01}}{\rho}
\right)\bar m\right\}\right],
\eeq
where we have used eqns. \eqref{var_Lagrangian}, \eqref{sigma_exp} and \eqref{connection_delta} in the first line and eqns. \eqref{NP_eqna} in the second line.  The first term in the above expression is exact but not others.
However, we show that the contribution of the scalar field is such that the symplectic current is exact and one may obtain a boundary symplectic current.

The symplectic current for the real scalar field is given by, 
$J_M(\delta_1,\delta_2)=2\,\delta_{[1}\varphi~\delta_{2]}\,{}\star d\varphi$.
The symplectic current on the hypersurface $\Delta$ can be obtained as
\begin{equation}
\underleftarrow{J_M}(\delta_1,\delta_2)=2\delta_{[1}\varphi~\delta_{2]}
(D\varphi~ n\wedge im\wedge\bar{m}),
\end{equation}
where $D=l^{a}\nabla_{a}$. The boundary condition on the scalar field implies 
$D\varphi=-2\rho \,\varphi$ and hence, we get 
that 
\begin{eqnarray}
\underleftarrow{J_M}(\delta_1,\delta_2)&=&4\delta_{[1}\varphi~\delta_{2]}
(-\varphi\,\rho~n\wedge im\wedge\bar{m})\nn
&=&
-d\left\{\delta_{[1}\varphi^2~\delta_{2]}~^2\epsilon\right\}+\delta_{
[1}~^2\epsilon~\delta_{2]}
\left(\frac{{ T}_{11}}{\rho}n-\frac{{ T}_{14}}{\rho}m-\frac{{ T}_{13}}{\rho}\bar m\right)\nn
\end{eqnarray}
The combined expression obtained by using the Einstein field equations, is then given by:
\begin{equation}
\underleftarrow{J_{M+G}}(\delta_1,\delta_2) \stackrel{\Delta}{=}-\frac{1}{4\pi 
G}\left\{d\left(\delta_{[1}~^2{\epsilon}~
\delta_{2]}\log{\rho}\right)\right\}-d\,\left\{\delta_{[1}\varphi^2~\delta_{2]}
~^2\epsilon\right\}.
\end{equation}
It follows that the hypersurface independent symplectic structure is given by:
\begin{eqnarray}
\Omega(\delta_{1}, 
\delta_{2})&=&\int_{\mathcal{M}}J_{M+G}(\delta_1,\delta_2)-\int_{S_\Delta}j(\delta_1,\delta_2)\nonumber\\
&=&-\frac{1}{8\pi G}
\int_{\mathcal{M}}\delta_{[1}\Sigma^{IJ}\wedge~\delta_{2]}A_{IJ}+2\int_{\mathcal
{M}}\delta_{[1}\varphi~\delta_{2]}
(\star d\varphi)
+\frac{1}{4\pi 
G}\int_{S_{\Delta}}\left\{\delta_{[1}~^2{\epsilon}~\delta_{2]}\log{\rho}
\right\}
+\int_{S_{\Delta}}\delta_{[1}\varphi^2~\delta_{2]}~^2\epsilon
\end{eqnarray}
In the next section, we shall use this expression to derive the first law of 
mechanics for the conformal Killing horizon.

\subsection*{Angular momentum as Hamiltonian}
Angular momentum is usually defined as a conserved charge for an axial Killing vector. However, in dynamic situations charges may not be conserved. One requires a way to define angular momentum which would remain valid in the time- independent case too. Consider that angular momentum to be the Hamiltonian function corresponding to a space-like rotational vector-field. If this vector field is also Killing, the Hamiltonian function will match with the angular momentum obtained as a conserved charge of the Killing vector field. In our case it turns out that there is a choice of such a vector field, as discussed. We will therefore define the angular momentum to be the Hamiltonian corresponding to the spacelike conformal Killing vector $\phi^a$.
We need to impose a few conditions on the fields to make a well defined 
Hamiltonian. These conditions are required since
the action of $\delta_{\phi}$ on some phase- space fields in not like $\lie_{\phi}$.  
First, we note the following equalities
\begin{eqnarray}
\lie_{l}\left(\frac{1}{4\pi G}\log{\rho}-\frac{1}{8\pi 
G}\log{\varphi}-\varphi^2\right)&=& \frac{1}{4\pi 
G}(2\rho+\epsilon+\bar{\epsilon})\\
\lie_{l}\left(\frac{~^2\epsilon}{\varphi}\right)&=&0
\end{eqnarray}
We assume that $\delta_\phi$ acts on $(2\rho+\epsilon+\bar{\epsilon})$ and 
$\left(\frac{~^2\epsilon}{\varphi}\right)$ like $\lie_\phi$. 
Moreover, since $\delta_{\phi}\lie_{l}(2\rho+\epsilon+\bar{\epsilon})=0$ it 
immediately implies that 
$\lie_{l}\delta_{\phi}(2\rho+\epsilon+\bar{\epsilon})=0$. Hence, one may choose 
the variables in such a way that 
 $\delta_{\phi}(2\rho+\epsilon+\bar{\epsilon})=0$  This 
implies that if we set $\delta_\phi\left(\frac{1}{4\pi G}\log{\rho}-\frac{1}{8\pi 
G}\log{\varphi}-\varphi^2\right)$=0 
at the initial cross-section, it remains zero everywhere on $\Delta$ and so,
\begin{eqnarray}\label{eqn_no1}
\frac{\delta_\phi\rho}{\rho}-8\pi 
G\varphi\delta_\phi\varphi-\frac{\delta_\phi\varphi}{2\varphi}=0
\end{eqnarray}
Another condition can be derived from the equation above
\begin{eqnarray}
\delta_\phi\left(\frac{~^2\epsilon}{\varphi}\right)=\frac{1}{\varphi}
\delta_\phi~^2\epsilon-~^2\epsilon\frac{1}{\varphi^2}\delta_\phi\varphi=0
\end{eqnarray}
The variations $\delta_\phi$ satisfy the following differential equations, which 
can be checked to be consistent with each other:
\begin{eqnarray}
\label{eqn_no2}
\lie_l\delta_\phi\varphi&=&-2\delta_l\rho\varphi-2\rho\delta_\phi\varphi\\
\lie_l\delta_\phi~^2\epsilon&=&-2\delta_\phi\rho~^2\epsilon-2\rho\delta_\phi~^2\epsilon
\end{eqnarray}
Putting condition $\eqref{eqn_no1}$ in $\eqref{eqn_no2}$, we get
\begin{equation}
\delta_\phi\varphi=C(\theta,\phi)\exp\left[{-\int\left(16\pi 
G\varphi^2+3\right)\rho dv}\right],
\end{equation}
where $C(\theta,\phi)$, is a constant of integration. If we choose this 
constant $C(\theta,\phi)=0$, it immediately implies that 
$\delta_{\phi}\varphi=0=\delta_{\phi}{}^{2}\epsilon.$ With the choice of $\delta_\phi$ 
only the bulk symplectic structure survives,
\beq
\Omega(\delta,\delta_\phi)=-\frac{1}{16\pi 
G}\int_{S_{\Delta}}\left[(\phi.A_{IJ})\delta\Sigma^{IJ}-(\phi.\Sigma^{IJ})\wedge 
\delta A_{IJ}\right]+\int_{S_{\Delta}}\delta\varphi~(\phi\cdotp{}{\star} d\varphi)+\delta J_{\infty}
\eeq
Note that the matter field part in the above expression will not contribute. We have also assumed that at infinity the contribution to the symplectic structure is a total variation $\delta J_{\infty}$. It follows that
\beq
\Omega(\delta,\delta_\phi)&=&\delta H^{\phi}\nn
\nn
&=&-\frac{1}{8\pi G}\int_{S_\Delta}\left[(\bar\alpha+\beta)(\phi.\bar m)+(\alpha+\bar\beta)(\phi.m)\right]\delta(~^2\epsilon)+\frac{1}{8\pi G}\int_{S_\Delta}(\phi.~^2\epsilon)\wedge\delta\left[(\bar\alpha+\beta)(\bar m)+(\alpha+\bar\beta)(m)\right]+\delta J_{\infty}\nn
\nn
&=&-\delta\left[\frac{1}{8\pi G}\int_{S_\Delta}\left[(\bar\alpha+\beta)(\phi.\bar m)+(\alpha+\bar\beta)(\phi.m)\right]~^2\epsilon\right]+\delta J_{\infty}\nn
\nn
&=&-\delta\left[\frac{1}{8\pi G}\int_{S_\Delta}(\phi\cdot\omega)~^2\epsilon\right]+\delta J_{\infty}
\eeq
Now we define $J_\Delta^\phi=H^\phi-J_{\infty}$ to be the horizon angular momentum. This is consistent with the fact that the angular momentum at $\Delta$ is the difference of the contribution at infinity and the total angular momentum. Then \beq J_\Delta^\phi=-\frac{1}{8\pi G}\int_{S_\Delta}(\phi\cdot\omega)~^2\epsilon\eeq.
Note that while defining a temperature we used the conformally transformed connection $\tilde\omega$ (with the choice $\lie_l\log\Omega=\rho$). So ideally one should be defining an angular momentum to be $\frac{1}{8\pi G}\int_{S_\Delta}(\phi\cdot\tilde\omega)~^2\epsilon$. In the next few lines we will show that these two definitions are equivalent. The transformation of the pull-back of the connection on to to $\Delta$ has already been discussed. It follows that,

\beq
\int_{S_\Delta}(\phi\cdot\tilde\omega)~^2\epsilon=\int_{S_\Delta}(\phi\cdot\omega)~^2\epsilon+\int_{S_\Delta}\lie_{\phi}\log{\Omega}~^2\epsilon
\eeq
 It therefore follows that
\beq
\int_{S_\Delta}\lie_{\phi}\log{\Omega}~^2\epsilon=\int_{S_\Delta}g~^2\epsilon=-\frac{1}{2}\int_{S_\Delta}\left[d(\phi\cdot~^2\epsilon)+\phi\cdot d(~^2\epsilon)\right]
\eeq
Since $\phi^a$ is purely tangential to $\Delta$, it follows that the last integration is zero. Morever since we are integrating over a compact surface, the first integral is also zero. It therefore follows that.
\beq
\int_{S_\Delta}(\phi\cdot\tilde\omega)~^2\epsilon=\int_{S_\Delta}(\phi\cdot\omega)~^2\epsilon
\eeq

\subsection*{Hamiltonian evolution and the first law}
Given the symplectic structure, we can proceed to study the evolution of the 
system. We assume that there exists a vector
which gives the time evolution on the spacetime. Given this vector field, one 
can define a corresponding vector field on the phase- space 
which can be interpreted as the infinitesimal generator of time evolution in 
the covariant phase- space. We assume there is a vector field $t^a$ in space-time which generates time evolution. It then follows that $(t^a+\Omega_{\Delta}\phi^a)~\epsilon~[l^a]$, where $\Omega_{\Delta}$ is a constant on $\Delta$ but may vary over the space of solutions. Therefore $t^a=l^a-\Omega_{\Delta}\phi^a$ is a live vector field.
\begin{eqnarray}
\Omega(\delta,\delta_t)&=&-\frac{1}{16\pi 
G}\int_{S_{\Delta}}\left[(t.A_{IJ})\delta\Sigma^{IJ}-(t.\Sigma^{IJ})\wedge 
\delta A_{IJ}\right]
+\int_{S_{\Delta}}\delta\varphi~(t\cdotp{}{\star} d\varphi)\nn
\nn
&&\hspace{1cm}+\frac{1}{8\pi 
G}\int_{S_{\Delta}}\left(\delta~^2{\epsilon}~\delta_{t}\log{\rho}-\delta_t~^2
{\epsilon}
~\delta\log{\rho}\right)+\int_{S_{\Delta}}\frac{1}{2}(\delta\varphi^2~\delta_{t}
~^2\epsilon-\delta_t\varphi^2\delta~^2\epsilon)\nn
\end{eqnarray}
\begin{eqnarray}
\delta \tilde H_t=\Omega(\delta,\delta_t)&=&-\frac{1}{8\pi 
G}\int_{S_\Delta}(\rho+\epsilon+\bar{\epsilon})\delta~^2\epsilon-\Omega_{\Delta}\delta J^{\phi}_{\Delta}+
\frac{1}{8\pi G}\int_{S_\Delta}~^2\epsilon~(-2\delta\rho-8\pi G\,\delta\varphi D\varphi)+\delta E^{\infty}\\
\delta H_t&=&-\frac{1}{8\pi G}\int_{S_\Delta}(2\rho+\epsilon+
\bar{\epsilon}){\delta}~^2\epsilon-\Omega_{\Delta} \delta J_\Delta^\phi-\frac{1}{8\pi 
G}\int_{S_\Delta}\left[~^2\epsilon~({\delta}\rho+8\pi G\,{\delta}\varphi 
D\varphi)\right]+\delta E_\infty
\end{eqnarray}
Where we have redefined our Hamiltonian $H_t=\tilde{H}_t+\int_{S_\Delta}\rho~^2\epsilon$. This redefination is possible since the definition of the Hamiltonian is ambiguous upto a total variation. Further, as expected $\Omega(\delta_{t},\delta_t)=0$. Next we define, $E^t_{\Delta}=E_\infty-H_t,$ as the horizon energy. It is clear from above that for $\rho\rightarrow 0$ (i.e in the isolated horizon limit) it matches with the definition in \cite{Ashtekar:2002ag,Ashtekar:2003hk} if asymptotics is flat and $E_\infty=E_{ADM}$. It therefore follows that:
\beq
-{\delta}E^t_\Delta =-\frac{1}{8\pi G}\int_{S_\Delta}(2\rho+\epsilon+
\bar{\epsilon}){\delta}~^2\epsilon-\Omega_{\Delta}\delta J^\phi_\Delta-\frac{1}{8\pi 
G}\int_{S_\Delta}\left[~^2\epsilon~({\delta}\rho+8\pi G\,{\delta}\varphi 
D\varphi)\right].
\eeq

To recover the the 
more familiar form of first law known for a dynamical situation, we  assume 
there is a vector field $\tilde{\delta}$ on phase space which acts only on the 
fields on
$\Delta$ (and not in the bulk) such that it's action on the boundary variables 
is to evolve the boundary 
fields along the affine parameter $v$ (it may be interpreted to be a time 
evolution, like $\lie$). Now demanding that $\tilde{\delta}$ to be 
Hamiltonian would give an integrability condition which also ensures that 
$\delta_t$ is Hamiltonian. So 
one can calculate $\Omega(\tilde{\delta},\delta_t):=\tilde{\delta}H_{t}$,
%
%
which can be written in the following form,
\begin{equation}\label{firstlaw}
\dot{E^t_\Delta}=\,\frac{1}{8\pi G}\left(2\rho+\epsilon+\bar{\epsilon}\right)\dot{A}+\Omega_{\Delta}\dot{J_\Delta^\phi}+\frac{1}{8\pi 
G}\int_{S_\Delta}\left[~^2\epsilon~(\dot\rho+8\pi 
G\,\dot\varphi D\varphi)\right]
\end{equation}
where dots imply changes in the variables produced by the action of $\tilde{\delta}$.
%
%
Note that if $\tilde{\delta} =\lie_{t}$, then, $\tilde{\delta}\varphi D\varphi$ gives the expression $T_{ab}t^{a}l^{b}$.
Equation \eqref{firstlaw} is the form of evolution for the conformal Killing horizons. The first term in the above expression is the usual $TdS$ term while the second term is a flux term which takes into account the non-zero matter flux across $\Delta$.

\section{Discussions}

In this paper, we have developed the formalism of rotating quasi-local 
conformal Killing horizon. We have constructed the phase- space, the symplectic structure on the space of solutions and showed that 
a form of first law arises which can be written in the differential form. More specifically,
the first law can be written as $dU=TdS +\Omega_{H} dJ+$ flux terms, where the flux arises due to
the matter fields falling in through the horizon. The phase- space of the above construction consists of all those solutions which, on the horizon, satisfy the boundary conditions of a CKH as mentioned in the section II. These solutions, which constitute the phase- space, are in a conformal class. In other words, the space of solutions constitute points which are such that their geometrical quantities are covariant under a specific conformal transformation. For example, the connections induced on $\Delta$, given by $\omega$ and $\tilde\omega$ ( see eqn.  \eqref{conftr_omega}) and the surface gravity obtained from it are in the same equivalence class.
As a result of this construction, the first law holds good for all solutions in phase- space which are related by a specific conformal transformation.

Some comments on the angular momentum are as follows. Generally, angular momentum is defined to be the conserved charge corresponding to an axial Killing vector. However, in absence of such a Killing vector, one may define angular momentum to be a Hamiltonian function corresponding to an angular vector field. This is precisely what we have done here. However, in doing so, one must keep in mind that a conformal Killing vector field of a generic sphere may also be considered as a conformal Killing vector of a round metric. The conformal transformations of a two- sphere is the M\"{o}bius group $SL(2, \mathrm{C})/\mathsf{Z}_{2}$ . Also, since the homomorphism $SL(2,\mathrm{C})$ to the restricted Lorentz group $SO^{+}(1,3)$ is surjective
with kernel $1\, \mbox{and}\,  -1$, it is possible to study the relevant generators 
on the two- sphere from the generators of the restricted Lorentz group. There are three types of restricted Lorentz transformations: rotations in spacelike 2- plane, boosts in a timelike two plane and null rotations. Examples of vector fields generating such transformations are given by $(y\partial_{z}-z\partial_{y})$, $(x\partial_{t}+t\partial_{x})$ and $(y\partial_{t}+t\partial_{y}) + (y\partial_{x}-x\partial_{y})$ respectively.  Out of these, the only non- trivial vector field on the sphere is the rotation generator. This arises as follows: all the (conformal ) Killing vectors on the sphere can be written as proportional to $\epsilon^{ab}\partial_{a}w$ and hence $w$ is constant on the orbits of this vector field. Further, the boosts and null rotations have zeros on the sphere and hence the only non- trivial vector field is the rotation generator. We have also shown that one can obtain a conformal rotation vector field whose Hamiltonian is the angular momentum. Further, in case $\rho\rightarrow 0$, the expression for angular momentum exactly matches with those obtained for isolated horizons (which, on $\Delta$, is again exactly same as the Komar integral).

Given  a form of the first law, it may be compared to the first law of 
thermodynamics. However, since the horizon is growing, it
describes a non- equilibrium situation. It is not surprising that one gets a differential form of first law since in the case of conformal Killing 
horizons, the horizon identifies a 'time' and hence a meaningful notion of time translation arises.
This leads to a definite identification of temperature and entropy. Also, the first law contains 
the expansion $\theta=-2\rho$ and this reflects the fact that the horizon is expanding. This is also natural
from the perspective of non- equilibrium thermodynamics since quite generally, the first law in these cases must also include how the evolution takes place. Unlike the equilibrium change, where only the end points of evolution matters and not the path, here, the system remembers the path through which the non- equilibrium evolution has taken place. The parameter $\theta$ is precisely that information.
One may also enquire if that entropy can arise from some counting
of microstates. 
In case of isolated horizons, the boundary symplectic structure has a natural interpretation of being the symplectic structure of a field theory
residing on the boundary. A quantization of the boundary theory
 therefore provides a microscopic 
description of the entropy of the 
isolated horizon. The situation is similar here and hence it remains to see if such an interpretation may be given for conformal Killing horizons too. 

It is useful to 
compare and contrast the present analysis with the formalism of dynamical horizons. Compared to dynamical horizons, our results 
are restrictive. Dynamical horizons capture the dynamics of evolving horizons in greater generality
by considering non- zero shear (which in our case is zero since we are not interested at the present 
to include gravitational flux) and that the laws of evolution are integrated versions and hence capture global information of evolution. 
The black hole evolutions equations in ref [18] are based on $3+1$ decomposition of metric variables.
In comparison, our derivation of laws of mechanics for CKHs is based on phase- space analysis in the first order formalism. 
Moreover, the first law is derived as a Hamiltonian and is written in the differential form. This is essentially due to the 
fact that for the restricted class of evolving horizons that
we construct, a zeroth law exists. Our generalisation shows that it may be possible to address the questions regarding classical 
evolution of black holes in full generality using the phase- space formalism. Such a generalisation may also pave the way to 
associate the boundary symplectic structure with a suitable field theory leading to some understanding of entropy in 
these non- equilibrium contexts.

\appendix
\section{Proof that $2\rho+\kappa_l$ is constant on $\Delta$}
In this section we give a direct proof that the quantity $2\rho+\kappa_l$ is constant on $\Delta$. The proof closely follows the proof in \cite{Bardeen:1973gs}, except now we have a conformal Killing vector rather than a Killing vector. We conclude that unlike the proof in \cite{DyerHonig,Suldyer} the dominant energy condition has to necessarily hold 
in the conformally transformed spacetime rather than the physical spacetime.
If $\xi$ is a conformal Killing vector such that $\lie_\xi g_{ab}=2\phi g_{ab}$, then following identity holds,
\beq\label{CKI}
\nabla_c\nabla_a\xi_b=\nabla_c\phi~g_{ab}+\nabla_a\phi~g_{cb}+\nabla_b\phi~g_{ac}+R_{dcab}\xi^d
\eeq

Now suppose $\xi$ is a null hypersurface forming vector. Choose a null tetrad such that the NP $l^a$ is aligned along $\xi^a$. Then it follows that $\kappa_{NP}=\sigma=0$, $\rho-\bar\rho=0$ and $\phi=-\rho$. The accleration of the null vector is given by,
\beq\label{acc}
l^a\nabla_al^b&=&\kappa_l l^b\nn
-\kappa_l&=&n^bl^a\nabla_a l_b
\eeq
where $n$ is normalized such that $l.n=-1$ choose the complex terads $m^a,\bar m^a$ such that they are tangent to the space of generators $l^a$. From \eqref{acc} it follows that
\beq
-\delta\kappa_l=(m^c\nabla_cn^b)(l^a\nabla_al_b)+(m^c\nabla_c l^a)(n^b\nabla_al_b)+n^bl^am^c\nabla_c \nabla_a l_b
\eeq
Consider the first term on the right hand side
\beq
(m^c\nabla_cn^b)(l^a\nabla_al_b)&=&m^c\nabla_c n^b(\kappa_l l_b)=-\kappa_l m^c\nabla_cl_b=\kappa_l(\bar\alpha+\beta)
\eeq
The second term gives
\beq
\left((\bar\alpha+\beta)l^a-\rho m^a\right)\left(\kappa_l n_a+\gamma+\bar\gamma l_a-(\bar\alpha+\beta)\bar m^a-(\alpha+\bar\beta)m^a\right)=-\kappa_l(\bar\alpha+\beta)+\rho(\bar\alpha+\beta)
\eeq
Using identity \eqref{CKI} the third term can be written as,
\beq
\delta\rho+R_{dcab}l^d m^c l^a n^b
\eeq
If one assumes that NP $\Psi_1=0$, then 
\beq
-\delta\kappa_l=\delta\rho+\rho(\bar\alpha+\beta)+\frac{R_{13}}{2}
\eeq
Using the Ricci identity $\delta\rho=\rho(\bar\alpha+\beta)+\Phi_{10}$ the above gives $\delta(\kappa+2\rho)=0$. Similarly one can shoe that $\bar\delta(\kappa+2\rho)=0$.
To show that the quantity $\kappa+2\rho$, we compute the following
\beq
-D\kappa_l&=&(l^c\nabla_cn^b)(l^a\nabla_al_b)+(l^c\nabla_c l^a)(n^b\nabla_al_b)+n^bl^al^c\nabla_c \nabla_a l_b\\
\eeq
In equation \eqref{CKI} the first three terms give $2D\rho$ while the fourt term vanishes due to antisymmetry of $R_{dcab}$ under exchange $d\leftrightarrow c$.
\beq
-D\kappa_l&=&n^bl^al^c\nabla_c \nabla_a l_b=2D\rho
\eeq
which implies $D(\kappa_l+2\rho)=0$

\section{Evolution of Angular Momentum}
In this section we give a computation of the evolution of angular momentum along the cross-sections of $\Delta$. Unlike an isolated horizon case the angular momentum is not conserved. Hence $\lie_t$ of the angular is non-zero.  
The condition that $\phi^a$ commutes with $l^a$ imposes certain restrictions on $\phi^a$.
Let us assume that the space-like conformal Killing vector can be written as $\phi=Am+\bar{A}\bar {m}$. The condition that it commutes with $l^a$ imposes the following restrictions,
\beq
[l,\phi]=DA~m+A[l,m]+c.c=0
\eeq
which implies
\beq
DA~m-A(\bar\alpha+\beta-\bar\pi)+A(\rho+\epsilon-\bar\epsilon)~m+c.c=0
\eeq
To find relations for $\phi$ to be a conformal Killind vector on $\Delta$. We consider the conformal Killing equation for $\phi$.
\beq
\nabla_{(a} \phi_{b)}=\delta A~\bar{m}_{(a} m_{b)}+\bar\delta A~{m}_{(a} m_{b)}-A(\rho+\epsilon-\bar\epsilon)~{n}_{(a} m_{b)}-DA~{n}_{(a} m_{b)}+A(\beta-\alpha)~\bar{m}_{(a} m_{b)}
\eeq
It immediately follows that for $\phi$ to be a conformal Killing vector, the following relations should hold,
\beq
DA+A(\rho+\epsilon-\bar\epsilon)&=&0\\
\bar \delta A&=&0\\
A(\bar\alpha+\beta-\bar\pi)+c.c&=&0
\eeq

We now show how the angular momentum evolves on $\Delta$. We start with the expression for the angular momentum

\beq\label{ang_momentum_app}
J^{\phi}_{\Delta}=\int_{S_\Delta}\left[A(\bar\alpha+\beta)+\bar A(\alpha+\bar\beta)\right]~^2\epsilon
\eeq
To calculate $\lie_lJ^{\phi}_{\Delta}$ we note the following Ricci identities,
\beq
D\alpha-\bar\delta\epsilon&=&\alpha(\rho+\bar\epsilon-2\epsilon)-\bar\beta\epsilon+\pi(\epsilon+\rho)+{\Phi}_{01}\nn
\nn
D\bar\alpha-\delta\bar\epsilon&=&\bar\alpha(\bar\rho+\epsilon-2\bar\epsilon)-\beta\bar\epsilon+\bar\pi(\bar\epsilon+\bar\rho)+{\Phi}_{10}\\
\nn
D\beta-\delta\epsilon&=&\bar\rho(\bar\alpha+\beta)+\bar\pi\bar\rho+\bar\pi(\epsilon+\bar\epsilon)-(\bar\alpha+\beta)2\bar\epsilon+\Psi_1
\eeq
\\
Combining the above equations we get the following relations,
\beq
AD(\bar\alpha+\beta)+\bar AD(\alpha+\bar\beta)-\phi^a\nabla_a(\epsilon+\bar\epsilon)&=&\rho(A(\bar\alpha+\beta)+\bar A(\alpha+\bar\beta))+\rho(A\bar\pi+\bar A\pi)+(\epsilon+\bar\epsilon)(A\bar\pi+\bar A\pi)\nn
&&-A(\bar\alpha+\beta)2\bar\epsilon-\bar A(\alpha+\bar\beta)2\epsilon+A\Psi_1+\bar A\bar\Psi_1+A\Phi_{10}+\bar A\Phi_{01}\\
\nn
AD(\bar\alpha+\beta)+\bar AD(\alpha+\bar\beta)-\phi^a\nabla_a(\epsilon+\bar\epsilon)&=&2\rho(A(\bar\alpha+\beta)+\bar A(\alpha+\bar\beta))+(\epsilon+\bar\epsilon)(A(\bar\alpha+\beta)+\bar A(\alpha+\bar\beta))\nn
&&-A(\bar\alpha+\beta)2\bar\epsilon-\bar A(\alpha+\bar\beta)2\epsilon+A\Psi_1+\bar A\bar\Psi_1+A\Phi_{10}+\bar A\Phi_{01}\\
\nn
AD(\bar\alpha+\beta)+\bar AD(\alpha+\bar\beta)-\phi^a\nabla_a(\epsilon+\bar\epsilon)&=&2\rho(A(\bar\alpha+\beta)+\bar A(\alpha+\bar\beta))+(\epsilon-\bar\epsilon)A(\bar\alpha+\beta)+(-\epsilon+\bar\epsilon)\bar A(\alpha+\bar\beta)\nn
&&A\Psi_1+\bar A\bar\Psi_1+A\Phi_{10}+\bar A\Phi_{01}\\
\nn
AD(\bar\alpha+\beta)+\bar AD(\alpha+\bar\beta)-\phi^a\nabla_a(\epsilon+\bar\epsilon)&=&2\rho(A(\bar\alpha+\beta)+\bar A(\alpha+\bar\beta))+(-DA-\rho A)(\bar\alpha+\beta)+(-D\bar A-\rho \bar A)(\alpha+\bar\beta)\nn
&&A\Psi_1+\bar A\bar\Psi_1+A\Phi_{10}+\bar A\Phi_{01}
\eeq
Then it follows that,
\beq
D\left[A(\bar\alpha+\beta)+\bar A(\alpha+\bar\beta)\right]&=&\lie_\phi(\epsilon+\bar\epsilon)+\rho(A(\bar\alpha+\beta)+\bar A(\alpha+\bar\beta))+A\Psi_1+\bar A\bar\Psi_1+A\Phi_{10}+\bar A\Phi_{01}
\eeq
Hence the angular momentum given in eqn. \eqref{ang_momentum_app} evolves like,(note that $\lie_\phi J_\Delta^\phi=0$)
\beq
\lie_l J_\Delta^\phi&=&\int_{S_\Delta}\left[\lie_\phi(\epsilon+\bar\epsilon)-\rho(A(\bar\alpha+\beta)+\bar A(\alpha+\bar\beta))+A\Psi_1+\bar A\bar\Psi_1+A\Phi_{10}+\bar A\Phi_{01}\right]~^2\epsilon\nn
&=&\int_{S_\Delta}\left[\lie_\phi(\epsilon+\bar\epsilon)-\lie_\phi\rho+2A\Phi_{10}+2\bar A\Phi_{01}\right]~^2\epsilon\\
&=&\int_{S_\Delta}\left[-3\lie_\phi\rho+T_{ab}l^a\phi^b\right]~^2\epsilon
\eeq

\subsection*{Acknowedgements}
The authors acknowledge the discussions with Amit Ghosh and Palash Baran Pal. The authors also thank the anonymous referee for suggestions that made the presentation better. AC is partially supported through the UGC- BSR start-up grant vide their letter no. F.20-1(30)/2013(BSR)/3082. AG is supported by Department of Atomic-Energy, Govt. Of India. AG would also like to thank Central University of Himachal Pradesh, for the warm hospitality, where part of this work was done.

\end{document}